\documentclass[12pt]{article}

\usepackage{epsfig}

\newcommand{\lapproxeq}{\lower .7ex\hbox{$\;\stackrel{\textstyle  
<}{\sim}\;$}} 
\newcommand{\gapproxeq}{\lower .7ex\hbox{$\;\stackrel{\textstyle  
>}{\sim}\;$}} 
\newcommand{\stackdown}[2]{\lower 1.4ex\hbox{$\;\stackrel{\textstyle{#1}}  
{\scriptstyle{#2}}\;$}}
\newcommand{\beq}{\begin{equation}} 
\newcommand{\eeq}{\end{equation}} 
\newcommand{\bea}{\begin{eqnarray}} 
\newcommand{\eea}{\end{eqnarray}}

\def\(#1{ ^{(#1)} }

\def\beq{\begin{equation}}
\def\eeq{\end{equation}}

\newcommand{\form}[1]{(\ref{#1})}

\makeatletter 
\def\slash{\@ifnextchar[{\fmsl@sh}{\fmsl@sh[0mu]}} 
\def\fmsl@sh[#1]#2{%
  \mathchoice 
    {\@fmsl@sh\displaystyle{#1}{#2}}%
    {\@fmsl@sh\textstyle{#1}{#2}}%
    {\@fmsl@sh\scriptstyle{#1}{#2}}%
    {\@fmsl@sh\scriptscriptstyle{#1}{#2}}} 
\def\@fmsl@sh#1#2#3{\m@th\ooalign{$\hfil#1\mkern#2/\hfil$\crcr$#1#3$}} 
\makeatother 

\def\beq{\begin{equation}}
\def\eeq{\end{equation}}

\catcode`\@=11 

\def\lsim{\mathrel{\mathpalette\@versim<}}
\def\gsim{\mathrel{\mathpalette\@versim>}}
\def\@versim#1#2{\vcenter{\offinterlineskip
    \ialign{$\m@th#1\hfil##\hfil$\crcr#2\crcr\sim\crcr } }}

\def\t1{{\tilde 1}}

\def\slash#1{#1\hskip-6pt/\hskip6pt}

\def\to{\rightarrow}

\begin{document}

\begin{titlepage} 

\begin{flushright}
gr-qc/0407089\\
CERN--PH--TH/2004--134\\
ACT-05-04, MIFP-04-14
\end{flushright}
\vspace{0.4cm}

\begin{centering} 

{\Large \bf Brany Liouville Inflation } 

\vspace{0.8cm}

{\large
{\bf J.~Ellis$^a$}, {\bf N.E.~Mavromatos$^{b}$},
{\bf D.~V.~Nanopoulos$^c$}} \\
{}\
and \\
{}\
{\large
{\bf A.~Sakharov$^{a,d}$}
}

\vspace{0.8cm} 

$^a$ {\it Theory Division, Physics Department, CERN, CH-1211 Geneva 23, 
Switzerland}

$^b$ {\it King's College London, University of London, Department of Physics, 
Strand WC2R 2LS, London, U.K.} 

$^c$ {\it George P. and Cynthia W. Mitchell Institute for Fundamental 
Physics, Texas A\&M
University,\\ College Station, TX 77843, USA; \\
Astroparticle Physics Group, Houston
Advanced Research Center (HARC),
Mitchell Campus,
Woodlands, TX~77381, USA; \\
Academy of Athens,
Academy of Athens,
Division of Natural Sciences, 28~Panepistimiou Avenue, Athens 10679,
Greece}

$^d$ {\it Swiss Institute of Technology, ETH-Z\"urich, CH-8093 Z\"urich,
Switzerland}

\vspace{0.2cm}

{\bf Abstract}

\vspace{0.2cm}

\end{centering} 

We present a specific model for cosmological inflation driven by the
Liouville field in a non-critical supersymmetric string framework, in
which the departure from criticality is due to open strings stretched
between two moving Type-II 5-branes. We use WMAP and other data on
fluctuations in the cosmic microwave background to fix parameters of the
model, such as the relative separation and velocity of the 5-branes,
respecting also the constraints imposed by data on light propagation from
distant gamma-ray bursters. The model also suggests a small, relaxing
component in the present vacuum energy that may accommodate the breaking
of supersymmetry.

\vspace{0.4cm}
\begin{flushleft}
CERN--PH--TH/2004--134 \\
July 2004
\end{flushleft}

\end{titlepage}

\section{Introduction}

A plethora of recent astrophysical data, ranging from measurements of the
cosmic microwave background (CMB) with an unprecedented precision by
WMAP~\cite{wmap} to direct evidence for the acceleration of the Universe
from observations of high-redshift Type-Ia supernovae~\cite{supernovae},
support strongly two important characteristics of our observable
Universe~\cite{bellido}. (i) It seems to have undergone {\it cosmological
inflation}~\cite{inflation}, i.e., a phase with a near-exponentially
expanding scale factor in an approximately Robertson-Walker space-time
seems to have been an essential component of the early evolution of our
Universe, and (ii) $70\%$ of the Universe's present energy content does
not seem to be associated with any form of matter, and is termed {\it dark
energy}.

Inflationary dynamics is supported by the spatial flatness of the
Universe, and many of its aspects have been corroborated by the CMB
data~\cite{wmap}. In the standard field-theoretic implementation, an
inflationary epoch requires the presence of a scalar mode, the {\it
inflaton} field, whose nature is still unknown. Moreover, the precise
shape of its potential is not yet determined by the data from
WMAP~\cite{wmapinflaton} and other CMB experiments.

The flatness of the Universe, as well as other astrophysical observations,
requires the presence of dark energy to balance the energy budget of the
current Universe, as well as to explain the data on
supernovae~\cite{supernovae}. The dark energy may be either a strict {\it
cosmological constant} or a component of the vacuum energy that is
relaxing to zero, via some non-equilibrium process as in quintessence
models~\cite{carroll}, possibly following some excitation of our Universe
due to an initial catastrophic event. WMAP data have constrained the
present-day equation of state $p = w\rho$, where $p$ is the pressure and
$\rho$ the energy density of such a quintessence field, and found $ w
\lsim - 0.8$~\cite{wmap}. This is in agreement with the cosmological constant
model, which has $w = -1$, but does not require it.

Any non-trivial but constant vacuum energy density in
Friedman-Robertson-Walker cosmology would eventually dominate the
evolution of the Universe, causing it to re-enter an accelerating
inflationary phase. In the modern context of string theory~\cite{strings},
such a de-Sitter-like Universe is an unwelcome feature. This because it
implies the existence of an event horizon, which impedes the definition of
conventional asymptotic states, and thus an $S$-matrix~\cite{smatrix}.  
Since string theory is conventionally formulated in terms of $S$-matrix
elements, such a background would appear to be problematic. On the other
hand, relaxing quintessential scenarios, although suffering fine-tuning
drawbacks related to the shape of the scalar-mode potential, may allow the
definition of an asymptotic $S$-matrix, and hence may be easier to stomach
as solutions of some versions of string theory.

This situation has been discussed in the context of string
theory~\cite{HARC,papant}, in the modern context of brane
cosmology~\cite{branecosm}, and in a model involving colliding brane
worlds, one of which is considered as our observable
Universe~\cite{grav2}. Other colliding-world scenaria of the `ekpyrotic'
type have been discussed extensively in the recent
literature~\cite{ekpyrotic}, where it was suggested that an inflationary
phase was absent and unnecessary. However, this point of view may be
difficult to reconcile with the above-mentioned recent evidence for
inflation. Moreover, this approach has been criticized in a stringy
context~\cite{linde}, on the grounds that classical string equations of
motion (conformal invariance conditions)  do not lead to expanding
Universes but rather to contracting ones~\footnote{This last point has
been questioned, however, in~\cite{seiberg}, assuming the existence of a
hypothetical (non-perturbative) stringy phase transition.}.
 
A different point of view was advocated in~\cite{grav2}, where the
collision of the brane worlds has been viewed as a {\it non-equilibrium
stringy process}, formulated within a {\it non-critical} (Liouville)  
string theory~\cite{ddk,emn} and exploiting the identification of target
time with the zero mode of the Liouville mode~\cite{kogan1,emn,kogan}. In 
this
scenario, the catastrophic cosmic event due to the collision of the brane
worlds leads to a central charge deficit in the world-sheet $\sigma$ model
that describes the stringy excitations of our (brane) Universe. In the
context of the identification of the Liouville mode with target time, this
central deficit provides a starting-point for cosmic time.

An important consequence of this departure from critical string theory,
and thus from the standard conformal invariance conditions used
in~\cite{linde}, is the presence of an exponentially-expanding {\it
inflationary} phase for the four-dimensional cosmological scale factor.
Moreover, such models lead naturally to an asymptotic quintessential dark
energy component of the Universe which is currently relaxing to zero,
depending on the cosmic time as $1/t^2$, computed using logarithmic
conformal field theory~\cite{lcft} methods~\footnote{We note in passing
that, in the model of~\cite{grav2}, such a relaxing dark energy component
in the current era can be made compatible with standard
supersymmetry-breaking models, with the symmetry breaking scale in the TeV
range.}. The early inflationary and late accelerating phases of the
Universe are thus correlated and 
occur dynamically in such models, without the introduction of
extra scalar fields such as an inflaton or a scalar quintessence field.
Instead, in non-critical string theory, such inflationary phases may be
obtained~\cite{emninfl} as a result of identifying the target time with
the zero mode of the Liouville world-sheet $\sigma$-model
field~\cite{kogan1,emn}. The consistency of this procedure has been
checked in several models.

Here this approach is revisited in some detail for the colliding
brane-world scenario of~\cite{grav2}, which is improved to incorporate
space-time supersymmetry~\cite{emw}, as may be motivated by the stability
of the underlying brane configurations as well as particle-physics
considerations.  We then discuss the cosmological parameters of this
model, taking into account the motion of the D-branes. Since the collision
of branes, assumed to take place adiabatically, induces the inflationary
phase, we can constrain the recoil velocity of the branes after the
collision by CMB measurements. We also constrain the distance between the
branes and the string coupling in this scenario, exhibiting a region of
parameter space that is compatible also with limits on deviations from
naive Lorentz invariance in the propagation of high-energy photons from
astrophysical sources such as gamma-ray bursters
(GRBs)~\cite{nature,mitsou,wavelets}. We also discuss the prospects for
dark energy and supersymmetry breaking in this scenario.

\section{Inflation as a Liouville String $\sigma$ Model}

\subsection{Inflation from Generic Liouville String Models} 

Before presenting our specific model, we first discuss briefly how an
inflationary space-time may be derived generically as a consistent
background in a non-critical string theory~\cite{emninfl,mp}. The approach
could be applied to a wide range of non-critical string models, so we
summarize its general features~\cite{emninfl} before applying it to the
concrete brane model constructed in~\cite{emw}.

As discussed in~\cite{emninfl,grav2,mp}, a constant central-charge deficit
$Q^2$ in a stringy $\sigma$ model may be associated with an initial 
inflationary phase~\cite{aben}, with
\begin{equation}
\label{centraldeficit} 
Q^2 = 9 H^2 > 0~,
\end{equation}
where the Hubble parameter $H$ can be fixed in terms of other parameters
of the model. One may
consider various scenaria for such a departure from criticality. For example,
in the model of~\cite{grav2} this was due to a `catastrophic' cosmic 
event, namely the
collision of two brane worlds. In such a scenario, as we now review
briefly, it is possible to obtain an initial {\it
supercritical } central charge deficit, and hence a time-like Liouville
mode in the theory. For instance, in the specific colliding-brane model
of~\cite{grav}, $Q$ (and thus $H$) is proportional to the square of the
relative velocity of the colliding branes, $Q \propto u^2$ during the
inflationary era.  As is evident from (\ref{centraldeficit}) and discussed
in more detail below, in a phase of constant $Q$ one obtains an
inflationary de Sitter Universe.

However, cosmically catastrophic non-critical string scenaria, such as
that in~\cite{grav2}, allow in general for a time-dependent deficit
$Q^2(t)$ that relaxes to zero. This may occur in such a way that, although
during the inflationary era $Q^2$ is (for all practical purposes)
constant, as in (\ref{centraldeficit}), eventually $Q^2$ decreases with
time so that, at the present era, one obtains compatibility with a new
accelerating phase of the Universe. As already mentioned, such relaxing
quintessential scenaria~\cite{grav2,papant} have the advantage of
asymptotic states that can be defined properly as $t \to \infty$, as well
as a string scattering $S$-matrix~\footnote{Another string scenario for
inducing a de Sitter Universe envisages generating the inflation
space-time from string loops (dilaton tadpoles)~\cite{fischler}, but in
such models a string $S$-matrix cannot be properly defined.}.

The specific normalization in (\ref{centraldeficit})  is imposed because 
one may identify the time $t$ with
the zero mode of the Liouville field $-\varphi$ of the {\it supercritical
} $\sigma$ model. The minus sign may be understood both mathematically, as
due to properties of the Liouville mode, and physically by the requirement
of the relaxation of the deformation of the space-time following
the distortion induced by the recoil. With this identification, the 
general equation of motion for the couplings $\{ g_i \}$ of the 
$\sigma$-model 
background modes is~\cite{emn}:
\begin{equation}
{\ddot g}^i + Q{\dot g}^i 
= -\beta^i (g) = -{\cal G}^{ij} \partial C[g]/\partial g^j~,
\label{liouveq}
\end{equation}
where the dot denotes a derivative with respect to the Liouville
world-sheet zero mode $\varphi$, and ${\cal G}^{ij}$ is an inverse
Zamolodchikov metric in the space of string theory couplings $\{ g^i 
\}$~\cite{zam}. When
applied to scalar, inflaton-like, string modes, (\ref{liouveq})  would
yield standard field equations for scalar fields in de Sitter
(inflationary)  space-times, provided the normalization
(\ref{centraldeficit}) is valid, implying a `Hubble' expansion
parameter $H=-Q/3$~\footnote{The gradient-flow property of
the $\beta$ functions makes the analogy with the inflationary case even
more profound, with the running central charge $C[g]$~\cite{zam} playing
the r\^ole of the inflaton potential in conventional inflationary field
theory.}.  The minus sign in $Q=-3H$ is due to the fact that, as we
discuss below, one identifies the target time $t$ with the world-sheet
zero mode of $-\varphi$~\cite{emn}.

The relations (\ref{liouveq}) replace the conformal invariance conditions
$\beta^i = 0$ of the critical string theory, and express the conditions
necessary for the restoration of conformal invariance by the Liouville
mode~\cite{ddk}. Interpreting the latter as an extra target dimension, the
conditions (\ref{liouveq}) may also be viewed as conformal invariance
conditions of a {\it critical} $\sigma$ model in (D+1) target space-time
dimensions, where D is the target dimension of the non-critical $\sigma$
model before Liouville dressing.  In most Liouville approaches, one treats
the Liouville mode $\varphi$ and time $t$ as independent coordinates.  In
our approach~\cite{emn,papant,grav2}, however, we take a further step,
basing ourselves on dynamical arguments which restrict this extended
(D+1)-dimensional space-time to a hypersurface determined by the
identification $\varphi = -t$. This means that, as time flows, one is
restricted to this D-dimensional subspace of the full (D+1)-dimensional
Liouville space-time.

In the work of~\cite{grav2} which invoked a brane collision
as a source of departure from criticality, this restriction arose because
the potential between massive particles, in an effective field theory
context, was found to be proportional to ${\rm cosh}(t + \varphi)$, which
is minimized when $\varphi = -t$. However, the flow of the Liouville mode 
opposite
to that of target time may be given a deeper mathematical interpretation.  
It may be viewed as a consequence of a specific treatment of the area
constraint in non-critical (Liouville) $\sigma$ models~\cite{kogan1,emn},
which involves the evaluation of the Liouville-mode path integral via an
appropriate steepest-descent contour.  In this way, one obtains a
`breathing' world-sheet evolution, in which the world-sheet area starts
from a very large value (serving as an infrared cutoff), shrinks to a very
small one (serving as an ultraviolet cutoff), and then inflates again
towards very large values (returning to an infrared cutoff). Such a
situation may then be interpreted as a world-sheet `bounce' back to the
infrared, implying, following the reasoning of~\cite{coleman}, that the
physical flow of target time is opposite to that of the world-sheet scale
(Liouville zero mode).

We now become more specific.  We consider a non-critical $\sigma$ model in
metric ($G_{\mu\nu}$), antisymmetric tensor ($B_{\mu\nu}$), and dilaton
($\Phi$) backgrounds. These have the following ${\cal O}(\alpha ')$ 
$\beta$ functions, where $\alpha '$ is the Regge slope~\cite{tseytlin}:
\begin{eqnarray} 
&& \beta^G_{\mu\nu} = \alpha ' \left( R_{\mu\nu} + 2 \nabla_{\mu} 
\partial_{\nu} \Phi 
 - \frac{1}{4}H_{\mu\rho\sigma}H_{\nu}^{\rho\sigma}\right)~, \nonumber \\
&& \beta^B_{\mu\nu} = \alpha '\left(-\frac{1}{2}\nabla_{\rho} H^{\rho}_{\mu\nu} + 
H^{\rho}_{\mu\nu}\partial_{\rho} \Phi \right)~, \nonumber \\
&& {\tilde \beta}^\Phi = \beta^\Phi - \frac{1}{4}G^{\rho\sigma}\beta^G_{\rho\sigma} = 
\frac{1}{6}\left( C - 26 \right).
\label{bfunctions}
\end{eqnarray} 
The Greek indices are four-dimensional, including target-time
components $\mu, \nu, ...= 0,1,2,3$ on the D3-branes
of~\cite{grav2}, and $H_{\mu\nu\rho}= \partial_{[\mu}B_{\nu\rho]}$ is the
field strength.

We consider the following representation of the four-dimensional 
field strength in terms of a pseudoscalar (axion-like) field $b$: 
\begin{equation}
H_{\mu\nu\rho} = \epsilon_{\mu\nu\rho\sigma}\partial^\sigma b 
\label{axion}
\end{equation}
where $\epsilon_{\mu\nu\rho\sigma}$ is the four-dimensional antisymmetric
symbol. Next, we choose an axion background that is linear in the
time $t$~\cite{aben}:
\begin{equation} 
b = b(t) = \beta t~, \quad  \beta={\rm constant} ,
\label{axion2}
\end{equation}
which yields a constant field strength with spatial indices only: $H_{ijk}
= \epsilon_{ijk}\beta$, $H_{0jk}= 0$.  This implies that such a background
is a conformal solution of the full ${\cal O}(\alpha')$ $\beta$-function
for the four-dimensional antisymmetric tensor. We also consider a dilaton
background that is linear in the time $t$~\cite{aben}:
\begin{equation}
\Phi (t,X) = {\rm const} + ({\rm const})' t .
\label{constdil}
\end{equation}
This background does not contribute to the $\beta$ functions 
for the antisymmetric tensor and metric.

Suppose now that only the metric is a non-conformal background, due to 
some initial quantum fluctuation or catastrophic event, such as the 
collision of two branes discussed above and in~\cite{grav}, 
which results in an initial central charge deficit $Q^2$ 
(\ref{centraldeficit}) that is constant at early stages after the 
collision. Let 
\begin{equation} 
G_{ij} = e^{\kappa \varphi + Hct}\eta_{ij}~, \quad G_{00}=e^{\kappa '\varphi 
+ Hct}\eta_{00},
\label{metricinfl}
\end{equation}
where $t$ is the target time, $\varphi$ is the Liouville mode, 
$\eta_{\mu\nu}$ is the four-dimensional Minkowski metric, 
and $\kappa, \kappa '$ and $c$ are constants to be determined. 
As already discussed, the standard inflationary scenario in 
four-dimensional physics requires $Q = -3H$,
which partially stems from~\cite{kogan1,emn}, and
\begin{equation}
\varphi = -t.
\label{liouvtime}
\end{equation}
This latter restriction is imposed dynamically~\cite{grav,emn}
at the end of our computations. Initially, one should treat
$\varphi, t$ as independent target-space components. 

The Liouville dressing induces~\cite{ddk} $\sigma$-model terms of the form 
$\int_{\Sigma} R^{(2)} Q \varphi$, where $R^{(2)}$ is the world-sheet curvature.
Such terms provide non-trivial contributions to the dilaton background in 
the (D+1)-dimensional space-time $(\varphi,t,X^i)$:
\begin{equation}
\Phi (\varphi,t,X^i) = Q \,\varphi + ({\rm const})' t + {\rm const}.
\label{seventy}
\end{equation}
If we choose $({\rm const})'=Q$, (\ref{seventy}) implies a 
constant dilaton background.

We now consider the Liouville-dressing equations~\cite{ddk}
(\ref{liouveq})  for the $\beta$ functions of the metric and antisymmetric
tensor fields (\ref{bfunctions}). For a constant dilaton field, the
dilaton equation yields no independent information, apart from expressing
the dilaton $\beta$ function in terms of the central charge deficit as
usual. For the axion background (\ref{axion2}), only the metric yields a
non-trivial constraint (we work in units with $\alpha ' =1$ for
convenience):
\begin{equation} 
{\ddot G}_{ij} + Q{\dot G}_{ij} = -R_{ij} + \frac{1}{2}\beta^2 G_{ij},
\end{equation}
where the dot indicates differentiation with respect to the (world-sheet
zero mode of the) Liouville mode $\varphi$, and $R_{ij}$ is the
(non-vanishing) Ricci tensor of the (non-critical) $\sigma$ model with
coordinates $(t,{\vec x})$:  $R_{00}=0~, R_{ij}=\frac{c^2H^2}{2}e^{(\kappa
- \kappa ')\varphi}\eta_{ij}$. One should also take into account the
temporal ($t$) equation for the metric tensor (for the antisymmetric
backgrounds this is identically zero):
\begin{equation}
{\ddot G}_{00} + Q{\dot G}_{00} = -R_{00} = 0,
\label{tempgrav}
\end{equation}
where the vanishing of the Ricci tensor stems from the 
specific form of the background (\ref{metricinfl}).
We seek metric backgrounds of Robertson-Walker inflationary 
(de Sitter) form:
\begin{equation}
G_{00}=-1~, \quad G_{ij}=e^{2Ht}\eta_{ij}.
\label{desittermetric}
\end{equation}
Then, from (\ref{desittermetric}), (\ref{metricinfl}),
(\ref{constdil}) and (\ref{axion2}), and imposing
(\ref{liouvtime}) at the end, we observe that there indeed is a consistent
solution with:
\begin{equation}
Q = -3H = - \kappa ',~c=3,~\kappa = H,~\beta^2 = 5H^2,
\label{solution}
\end{equation}
corresponding to the conventional form of inflationary equations for
scalar fields.

\subsection{A Concrete Non-critical String Example: Colliding Branes}

\begin{figure}[htb]
\begin{center}
\epsfxsize=3in
\bigskip
\centerline{\epsffile{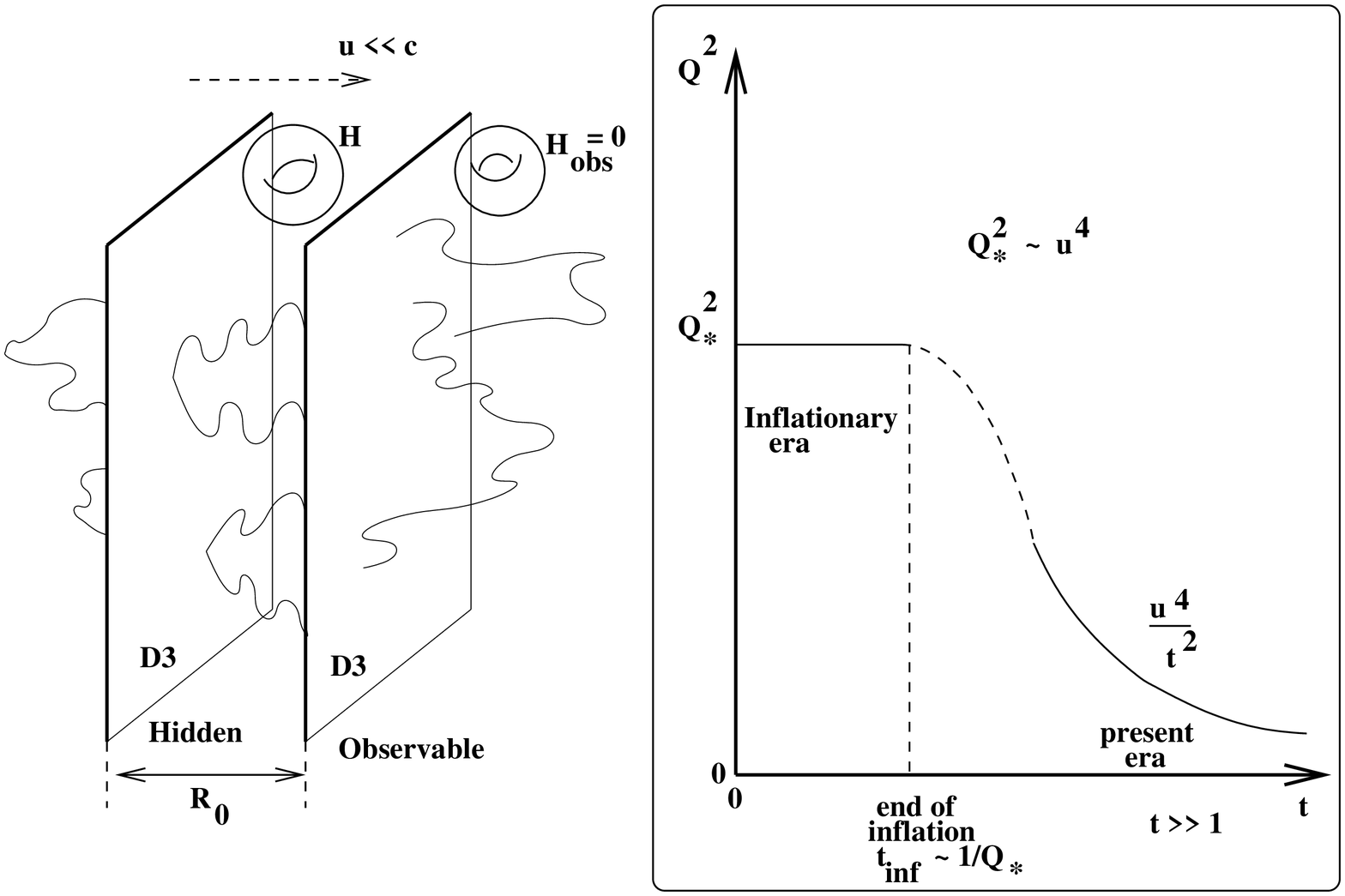}}
\caption{{\it A scenario 
in which the collision of two Type-II 5-branes provides 
inflation and a relaxation model for cosmological vacuum energy.\label{infla}}}
\end{center} 
\end{figure}

We now concentrate on one particular example of the previous general
scenario~\cite{emninfl}, in which the non-criticality is induced by the
collision of two branes, as seen in Fig.~\ref{infla}. We first discuss the
basic features of this scenario, and then proceed to demonstrate
explicitly the emergence of inflationary space-times from such situations.

Following~\cite{grav2}, we consider two 5-branes of Type-II string
theory, in which the extra two dimensions have been compactified on tori.
On one of the branes (assumed to be the hidden world), the torus is
magnetized with a field intensity ${\cal H}$.  Initially our world is
compactified on a normal torus, without a magnetic field, and the two
branes are assumed to be on a collision course with a small relative
velocity $v \ll 1$ in the bulk, as illustrated in Fig.~\ref{infla}. The 
collision produces a non-equilibrium
situation, which results in electric current transfer from the hidden
brane to the visible one.  This causes the (adiabatic) emergence of a
magnetic field in our world.

The instabilities associated with such magnetized-tori compactifications
are not a problem in the context of the cosmological scenario discussed
here.  In fact, as discussed in~\cite{grav2}, the collision may also
produce decompactification of the extra toroidal dimensions at a rate much
slower than any other rate in the problem. As discussed in~\cite{grav2},
this guarantees asymptotic equilibrium and a proper definition of an
$S$-matrix for the stringy excitations on the observable world.

The collision of the two branes implies, for a short period afterwards,
while the branes are at most a few string scales apart, the exchange of
open-string excitations stretching between the branes, whose ends are
attached on them. As argued in~\cite{grav2}, the exchanges of such pairs
of open strings in Type-II string theory
result in an excitation energy in the visible world.  The latter may be
estimated by computing the corresponding scattering amplitude of the two
branes, using string-theory world-sheet methods~\cite{bachas}: the time 
integral for the relevant potential yields the
scattering amplitude.  Such estimates involve the computation of
appropriate world-sheet annulus diagrams, due to the existence of open
string pairs in Type-II string theory. This implies the presence of `spin
factors' as proportionality constants in the scattering amplitudes, which
are expressed in terms of Jacobi $\Theta$ functions. For the small brane
velocities $v \ll 1$ we are considering here, the appropriate spin
structures start at {\it quartic order } in $v$, as a result of the
mathematical properties of the Jacobi functions~\cite{bachas}. This in
turn implies~\cite{grav2} that the resulting excitation energy on the
brane world is of order $V = {\cal O}(v^4)$, which may be thought of as an
initial (approximately constant) value of a {\it supercritical}
central-charge deficit for the non-critical $\sigma$ model that describes
stringy excitations in the observable world after the collision:
\begin{equation}
Q^2 = {\cal O}\left( v^4 \right) > 0.
\label{initialdeficit}
\end{equation} 
The supercriticality of the model is essential~\cite{aben} 
for a time-like signature of the Liouville mode, and hence its 
interpretation as target time.

At times long after the collision, the branes slow down and the central 
charge deficit is no
longer constant but relaxes with time $t$.  In the approach
of~\cite{grav2}, this relaxation has been computed by using world-sheet
logarithmic conformal field theory methods~\cite{lcft,kogan}, taking into
account recoil (in the bulk) of the observable-world brane and the
identification of target time with the (zero mode of the) Liouville field.  
This late-time varying deficit $Q^2(t)$ has been identified~\cite{grav2}
with a `quintessential' dark energy density component:
\begin{equation} 
\Lambda (t) \sim \frac{R^2({\cal H}^2 + v^2)^2}{t^2}
\left(\frac{M_s}{M_P}\right)^4M_P^4~,
\label{cosmoconst}
\end{equation} 
where $R$ is the compactification radius. This yields a present era dark
energy compatible in order of magnitude with the WMAP 
observations~\cite{wmap}.

The magnetic field ${\cal H}$ in the extra dimensions~\cite{grav2}
breaks target-space supersymmetry~\cite{bachas2}, due to
the fact that bosons and fermions on the brane worlds couple differently
to ${\cal H}$.  In our problem, where the magnetic field is turned on
adiabatically, the resulting mass difference between bosonic and fermionic
string excitations is found to be~\cite{grav2}:
\begin{equation} 
\Delta m^2_{\rm string} \sim 2{\cal H}{\rm cosh}\left(\epsilon \varphi + 
\epsilon t\right)\Sigma_{45},
\label{masssplit}
\end{equation} 
where $\Sigma_{45}$ is a standard spin operator in the plane of the torus,
and $\epsilon \to 0^+$ is the regulating parameter 
of the Heaviside operator $\Theta_\epsilon (t) = -i\int_{-\infty}^\infty 
\frac{d\omega}{\omega -i\epsilon}e^{i\omega t}$
appearing in the D-brane recoil formalism~\cite{kogan}.  
The dependence in (\ref{masssplit}) implies that the formalism selects 
dynamically a 
Liouville mode which flows opposite to the target time $\varphi = -t$,
as discussed earlier, as a result of minimization of the effective 
field-theoretic potential of the various stringy excitations. 
By choosing appropriately ${\cal H}$, we may arrange for the
supersymmetry-breaking scale to be of the order of a few TeV.  Such a
magnetic field contribution would be subdominant, compared with the
velocity contribution, in the expression
(\ref{cosmoconst}) for the present dark energy density. 

The model is capable in principle of reproducing naturally a value of the
present dark energy density (i.e., for $t \sim 10^{60}t_P$) that is
compatible with observations~\cite{supernovae,wmap}, provided one chooses
relatively large compactification radii $R \sim 10^{17}\ell_P \sim
10^{-18}$~m, which are common in modern string theories. In models where
the compactification involves higher-dimensional manifolds than tori, a
volume factor $R^{n}: n > 2$ is the number of extra dimensions, appears in
(\ref{cosmoconst}). In such cases, the compactification radii are
significantly smaller.

However, this (toy) model suffers from fine tuning, since the final
asymptotic value of the central charge deficit has been arranged to
vanish, by an appropriate choice of various constants appearing in the
problem~\cite{grav2}. This is required by the assumption that our
non-critical string system relaxes asymptotically in time to a critical
string.  In the complete model, the identification of the Liouville field
with target time~\cite{emn,kogan} would define the appropriate
renormalization-group trajectory, which hopefully would pick up the
appropriate asymptotic critical string state dynamically. This still
remains to be seen analytically in realistic models, although it has been
demonstrated numerically for some stringy models in~\cite{papant}.  
Nevertheless, the current toy example is sufficient to provide a
non-trivial, and physically relevant, concrete example of an inflationary
Universe in the context of Liouville strings.

\paragraph{}
\begin{figure}[tb]
\begin{center}
\includegraphics[width=4cm]{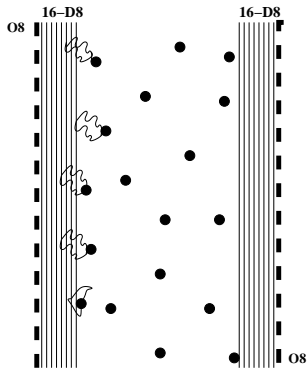}
\end{center}
\caption{\it A model for supersymmetric D-particle foam
consisting of two stacks each of sixteen parallel coincident D8-branes, 
with orientifold planes (thick dashed lines) attached to them.
The space does not extend beyond the orientifold planes.
The bulk region of ten-dimensional space in which the D8-branes 
are embedded is punctured by D0-particles (dark blobs). 
The two parallel stacks are sufficiently far from each other
that any Casimir contribution to the vacuum energy is negligible.
Open-string interactions between D0-particles and D8-branes
are also depicted (wavy lines). If the D0-particles are stationary,
there is zero vacuum energy on the D8-branes, and the configuration 
is a consistent supersymmetric string vacuum.}
\label{fig:nonchiral}
\end{figure}

This type of model can be extended to incorporate supersymmetry, as shown
in a recent paper~\cite{emw}. As illustrated in Fig.~\ref{fig:nonchiral},
this consists of two stacks of D8-branes with the same tension, separated
by a distance $R$. The transverse bulk space is restricted to lie between
two orientifold planes, and is populated by D-particles.  It was shown
in~\cite{emw} that, in the limit of static branes and D-particles, this
configuration constitutes a zero vacuum-energy supersymmetric ground state
of this brane theory. Bulk motion of either the D-branes or the
D-particles~\footnote{The latter could arise from recoil following
scattering with closed string states propagating in the bulk.} results in
non-zero vacuum energy and hence the breaking of target supersymmetry,
proportional to some power of the average (recoil) velocity squared, which
depends on the precise string model used to described the (open) stringy
matter excitations on the branes.

The colliding-brane scenario introduced earlier can be realized in this
framework by allowing (at least one of) the D-branes to move, keeping the
orientifold planes static. One may envisage a situation in which the two
branes collide, at a certain moment in time corresponding to the Big Bang
- a catastrophic cosmological event setting the beginning of observable
time - and then bounce back. The width of the bulk region is assumed to be
long enough that, after a sufficiently long time following the collision,
the excitation energy on the observable brane world - which corresponds to
the conformal charge deficit in a $\sigma$-model
framework~\cite{grav2,emw} - relaxes to tiny values. It is expected that a
ground state configuration will be achieved when the branes reach the
orientifold planes again (within stringy length uncertainties of order
$\ell_s=1/M_s$, the string scale). In this picture, since observable time
starts ticking after the collision, the question how the brane worlds
started to move is merely philosophical or metaphysical. The collision
results in a kind of phase transition, during which the system passes
through a non-equilibrium phase, in which one loses the conformal symmetry
of the stringy $\sigma$ model that describes perturbatively string
excitations on the branes. At long times after the collision, the central
charge deficit relaxes to zero~\cite{grav2}, indicating that the system
approaches equilibrium again. The dark energy observed today may be the
result of the fact that our world has not yet relaxed to this equilibrium
value. Since the asymptotic ground state configuration has static D-branes
and D-particles, and hence has zero vacuum energy as guaranteed by the
exact conformal field theory construction of~\cite{emw}, it avoids the
fine tuning problems in the model of~\cite{grav2}.

Sub-asymptotically, there are several contributions to the excitation
energy of our brane world in this picture. One comes from the interaction
of the brane world with nearby D-particles, i.e., those within distances
of order ${\cal O}(\ell_s)$, as a result of open strings stretched between
them.  These are of order $n_0{\cal V}$, where $n_0$ is the density of
D-particles on or near the brane world, and
\begin{equation}
{\cal V}_{D0-D8} \sim {\cal O}(u^2) f(R) 
\label{vd0d8}
\end{equation}
where $u$ the velocity of such a D-particle, which may in general be
different from the recoiling velocity $v$ of the colliding D-branes, and
$f(R)$ is an appropriate function of the distance between the D-particle
and the D-brane~\cite{gaber}, which is of order unity for distances of the
same order as the string scale. In the case of an isolated
D-particle/D8-brane system, there are also $u$-independent contributions
to ${\cal V}$, which, however, are cancelled in the construction
of~\cite{emw} by the addition of the other D8 branes and orientifold
planes.

The other contribution to the dark energy of our brane world comes from
the collision of the identical D-branes, which is of order ${\cal
O}(v^4)$, as mentioned above~\cite{grav2,emw}.  For a sufficiently dilute
gas of nearby D-particles we may assume that this latter contribution is
the dominant one. In this case, we may ignore the D-particle/D-brane
contributions (\ref{vd0d8}) to the vacuum energy, and hence apply the
previous considerations on inflation, based on the ${\cal O}(v^4)$ central
charge deficit, also in this scenario of colliding branes and D-particle
foam.

The presence of D-particles, which inevitably cross the D-branes in such a
picture, even if the D-particle defects are static initially, distorts
slighlty the inflationary metric on the observable brane world at early
times after the collision, during an era of approximately constant central
charge deficit, without leading to significant qualitative
changes~\cite{mp}. Moreover, the existence of D-particles on the branes
will affect the propagation of string matter on the branes, in the sense
of modifying their dispersion relations by inducing local curvature in
space-time, as a result of recoil following collisions with string matter.
However, it was argued in~\cite{emnequiv,emw} that only photons are
susceptible to such effects in this scenario, due to the specific gauge
properties of the membrane theory at hand.  The dispersion relations for
chiral matter particles, or in general fields on the D-branes that
transform non-trivially under the Standard Model gauge group, are
protected by special gauge symmetries in string theory, and as such are
not modified.  These specifically stringy reasons were outlined
in~\cite{emnequiv,emnequiv1}.

One may derive stringent limits on the possible modification of photon
dispersion relations using observations of gamma-ray bursters
(GRBs)~\cite{nature,mitsou,wavelets}. Writing the photon dispersion
relation as $E^2=p^2 + \xi \frac{p^3}{M_{\rm QG}}$, and restricting
ourselves to subluminal models with $\xi < 0$ as required by string theory
considerations~\cite{nature,emw}, we have found~\cite{wavelets}:
\begin{equation}
M_{\rm QG} \gsim 10^{16}~{\rm GeV}.
\label{qgscale}
\end{equation}
Limits on a possible modification of photon propagation stronger than 
(\ref{qgscale}) have been claimed in the literature,
but we do not consider them secure. Some are based on time-of-flight
analyses as proposed in~\cite{nature} using either a single
GRB~\cite{onegrb} - with unknown redshift and with no accounting for the
possible systematic uncertainty due to the possible energy-dependent souce
effect that was considered in ~\cite{mitsou,wavelets}, or a flare from an
Active Galactic Nucleus~\cite{mkr421} - where statistics is an issue as
well as a possible energy-dependent souce effect. There are other
constraints based on threshold analyses of absorption by the infrared (IR)
diffuse extragalactic background of TeV $\gamma$-rays emitted by
blazars~\cite{PM,steckerblazars,amelinoblasars} - which are vulnerable to
assumptions on the IR background and depend, in some cases, on assumptions
about the possible modifications of dispersion relations for other
particle species. However, these are not generic features of models of
quantum gravity, since violations of the equivalence principle appear, for
instance, in a stringy model of space-time
foam~\cite{emnequiv} related to the model used in this paper~\footnote{We 
note, also, that the strong limits
obtained from synchrotron radiation emission from the Crab
Nebula~\cite{jacobson,emnequiv1} apply to modifications of the dispersion
relation for electrons, which are absent in this models~\cite{emnequiv}.}.
For all the above reasons, we retain (\ref{qgscale}) as a
conservative and reliable limit for the purposes of the present work.

The relation between $M_{\rm QG}$ and the four-dimensional Planck scale
$M_P$ is a model-dependent issue. In models of D-particle foam, the
quantum gravity scale responsible for the modification of the dispersion
relation is the mass of the D-particle, $M_{\rm QG} = M_D = M_s/g_s$,
where $M_s$ is the string scale, and $g_s$ is the string coupling.

The string scale $M_s$ may or may not be the same as the four-dimensional
Planck scale $M_P \sim 10^{19}$ GeV. One scenario is to identify in our
case $M_s=M_P$, and then interpret the lower bound on $M_{\rm QG}$ found
in the analysis of~\cite{wavelets} as implying a real effect on photons,
with $M_{\rm QG}=M_D=M_s/g_s=M_P/g_s \sim 10^{16}~$GeV. This would imply
$g_s \le 10^{3}$.  On the other hand, one may identify $M_D$ with $M_P$,
in which case the results of~\cite{wavelets} may be interpreted only as
providing a sensitivity limit for quantum-gravity effects, three orders of
magnitude below the Planck scale. In this case, no information is obtained
on $g_s$ and $M_s$ separately from this experiment alone. 

\section{CMB Constraints on Brany Inflation} 

We now use WMAP data to set limits on the central charge
deficit $Q$, i.e., the recoil velocity $v$,
in our model of colliding branes, and also on the separation of the 
5-branes during inflation. We recall that, in our approach, 
the central charge deficit $Q^2$ of the D-particle space-time foam that is 
responsible for cosmological inflation is related to the Hubble 
expansion rate during inflation by $Q^2\simeq 9H^2$. Since the vacuum 
energy is dynamical in this scenario, we have the prospect of
a graceful exit from the inflationary epoch. In our Liouville string
model, inflation can be taken as ending when the colliding branes are
separated by a distance exceeding the string length $l_s$ by a few orders
of magnitude. If $t_{I}$ denotes the duration of inflation, we have:
\begin{equation}
vt_{I} = x\ell_s,
\label{equation} 
\end{equation} 
where $v$ is the relative velocity of the branes, and $x\gg 1$ is to be
determined below from observations of the CMB.  The amount of inflationary
expansion within a given timescale is usually parametrized in terms of the
number of $e$-foldings of the scale factor, denoted by $N$. This number
must be larger than about 60 (see~\cite{infl4} for details).  Assuming an
approximately constant relative velocity of the branes, the inflation
lasts for the right amount of $e$-foldings if the following condition is
satisfied:
\beq
\label{inflq3}
H_{I}\frac{x\ell_s}{v}=N\ge 60.
\eeq
A recent analysis~\cite{infl5} of WMAP data provides complementary
information about the energy scale during inflation, 
at the $2\sigma$ level:
\beq
\label{inflq4} 
\frac{H_{I}}{M_P}\le 1.48\times 10^{-5}.
\eeq
Combining \form{inflq4}, saturating the bound so as to fix ideas, 
with \form{inflq3}, we find
\beq
\label{inflq5}
\frac{v}{x}\simeq \frac{1.48}{N} \times 10^{-5}\frac{M_P}{M_s}.
\eeq
On the other hand, from our model above, we know that $H \ell_s \sim v^2$ 
as an order of magnitude estimate (up to factors pertaining to the volume of
compactified extra 
dimensions. In this section, we assume that the compactification radii are 
of order of the string scale $\ell_s$, otherwise such factors should be taken
properly into account).  On account of (\ref{inflq4}), then, we have 
\begin{equation}\label{velocity} 
v^2 \le 1.48 \times 10^{-5}\frac{M_P}{M_s}.
\end{equation}
One should consider the two cases mentioned in previous sections, 
namely that, either $M_P \simeq M_s$,
or $M_P = M_s/g_s$, with $g_s$ no less than $10^{-3}$.
In both cases we see that (\ref{velocity}) corresponds to comfortably 
non-relativistic relative motion of the branes, as we had implicitly assumed. 

Our non-critical string scenario realizes inflation dynamically, without
the explicit introduction of an extra inflaton field.  In our case, the
Hubble parameter $H$ depends on the relative velocity $v$ of the branes,
$H \propto v^2$. This is small and slowly-varying in cosmic time,
mimicking in essential respects the behaviour of a conventional scalar
inflaton field. In our stringy model, the recoil velocity $v$ corresponds
to a coupling of the underlying two-dimensional world-sheet $\sigma$
model, pertaining to logarithmic deformations~\cite{kogan,lcft}. As
described in~\cite{emn,kogan}, summation over world-sheet genera results
in the quantization of these deformations, inducing quantum fluctuations
of the `coupling' $v$ and thus also $H$.  Therefore, the constancy of the
dilaton field, based on the equality of the constants $Q$ and $({\rm
const})'$ in (\ref{seventy}), should be viewed only as a mean field
result. The summation over world-sheet genera, which corresponds to the
full quantum theory, leads to quantum fluctuations $\Delta Q \propto
\Delta v^2 = {\cal O}(g_s^2v^2)$, in $Q$~\cite{kogan}. These induce, in
turn, quantum fluctuations of the dilaton $\Phi$, $\Delta \Phi \sim
g_s^2v^2t$, with $t < t_{I}$, which should therefore be regarded as a
fully-fledged, canonically-normalized (in the so-called Einstein frame)
scalar quantum field, to be integrated over in a path integral of the
corresponding string field theory. The effective low-energy quantum theory
is thus a quantum field theory equivalent to a conventional slow-roll
inflation, with Einstein gravity at lowest order. This observation
justifies our subsequent application of the well-established behaviour of
quantum field fluctuations in a de Sitter background to the relative
velocity of the branes, that we use as a dynamical degree of freedom
driving inflation. We are therefore justified in using horizon-flow
parameters~\cite{infl6} to analyze the predictions of this Liouville
string model for inflation.

The horizon-flow functions are
generalizations of the usual slow-roll parameters~\cite{infl7}, and are
defined recursively as the logarithmic derivatives of the Hubble scale
with respect to the number of e-foldings:
\beq
\label{inflq6}
\epsilon_{i+1}=\frac{d\ln |\epsilon_i|}{dN}, \qquad i\ge 0, \qquad
\epsilon_0=\frac{H_I}{H}.
\eeq
where $H_I$ denotes an initial value of the Hubble parameter. 
In the case of a constant-horizon-$H$ inflation, as our case here, the spectral index for the power spectrum
is given in terms of horizon-flow parameters by the exact relation~\cite{infl6}: 
\beq
\label{inflq7}
n_S-1=-2\epsilon_1-\epsilon_2.
\eeq 
The weak energy condition (for a spatially flat universe) requires $\epsilon_1 > 0$, while inflation requires
$\epsilon_1 < 1$. 
Using (\ref{inflq6}) and the relation (\ref{inflq3}), 
and assuming that only $v$ varies with the number of e-foldings $N$, and not 
$H_I$ or the parameter $x$, 
one can easily see 
that the first three horizon-flow functions read:
\beq
\label{3flow}
\epsilon_0=\left(\frac{v_I}{v}\right)^2;\qquad \epsilon_1=2\frac{v}{x 
H_I\ell_s}=\frac{2}{N};
 \qquad \epsilon_2=-\frac{v}{x H_I\ell_s}=-\frac{1}{N}.
\eeq
from which we see that the weak energy condition for a spatially flat universe, requiring $\epsilon_1 > 0$, is
satisfied, and, that, 
due to (\ref{inflq3}), we also have $\epsilon_1 < 1$, as is the case in typical 
inflationary scenaria~\cite{infl6}.
One can then estimate the spectral index using \form{3flow} and 
\form{inflq7} as follows:
\beq
\label{inflq8}
n_S-1=-\frac{3 v}{H_Ix\ell_s} = -\frac{3}{N}.
\eeq    
The WMAP data~\cite{wmap}
are consistent
with a scale-invariant power spectrum~\cite{infl5} at the 1-$\sigma$ 
level~\footnote[1]{Similar sensitivity to the spectral index was reached
previously by combining data from Boomerang, Maxima-1 and
COBE~\cite{boomerang}.}, with 
\begin{equation} 
 n_s -1 = -4  \times 10^{-2} 
\label{wmapns}
\end{equation} 
From (\ref{wmapns}), (\ref{inflq8}) and (\ref{inflq5}) we obtain
\begin{equation}
\frac{v}{x}\frac{M_s}{M_P} \simeq 1.48 \times \frac{10^{-5}}{N} \simeq 1.97 \times 10^{-7} 
\label{vx}
\end{equation}
from which $N \simeq 75$. This result ensures that the horizon-flow
functions \form{3flow} are small, justifying {\it a posteriori} our
assumption of a slow-roll regime~\cite{infl6}. This is not surprising,
since the dependence of the Hubble rate on the relative velocity of
the branes is akin to a single-field (dilaton) slow-roll model of
inflation, for the reasons mentioned above.

We recall that, if we identify the four-dimensional quantum gravity scale
$M_P \sim 10^{19}$ GeV with $M_s$, and interpret the quantum gravity scale 
limit 
of~\cite{wavelets} (\ref{qgscale}) as referring to $M_s/g_s$ (the mass of 
the D-particles
in the foam), the string constant $g_s$ should be no less than $10^{-3}$
in the framework of inflationary Liouville string scenario.
From (\ref{velocity}), (\ref{vx}) then, we conclude that $g_s$ does not enter,
and 
\begin{equation}
x \ge 1.95 \times 10^{4}.
\label{final}
\end{equation}
This means that, at the end of inflation, the two recoiling branes
find themselves some $10^4\ell_s$ apart. On the other hand, if we assume 
$M_P/M_s =1/g_s$, in which case our GRB analysis does not yield a bound on 
$g_s$, a `reasonable' perturbative value $g_s \sim 10^{-2}$
in (\ref{velocity}), (\ref{vx}) would lead to 
\begin{equation}
x \ge 1.95 \times 10^{3}.
\label{final2}
\end{equation}
Either way, this simple analysis indicates that D-particle space-time foam
can accommodate an inflationary scenario that is consistent with the CMB 
data for a reasonable range of values of the string coupling.

\section{Summary and Outlook}

We have exhibited a specific brany scenario for cosmological inflation in
the general framework of non-critical string theory~\cite{emninfl}.  The
collision of two Type-II 5-branes~\cite{grav2} 
(or generalizations thereof to incorporate supersymmetric D-particle foam
in inflationary brany 
scenaria~\cite{emw})
causes a central-charge
deficit in the world-sheet $\sigma$ model related to their relative
velocity. The Hubble expansion rate during inflation is directly related
to this deficit, which is compensated by the Liouville field on the string
world sheet, whose zero mode is identified dynamically (up to a sign) with
cosmic time. Observations of the CMB by WMAP~\cite{wmap} and other
experiments provide constraints on inflationary
parameters~\cite{wmapinflaton} that may be interpreted as limits on the
relative velocity and separation of the colliding branes, for values of
the string coupling that are compatible with limits on the energy
dependence of photon propagation from astrophysical
sources~\cite{nature,mitsou,wavelets}. This brany scenario also provides
the possibility of a `quintessential' contribution to the present-day dark
energy that relaxes towards zero, as required for
supersymmetry~\cite{emw}. Meanwhile, in the presence of vacuum energy,
supersymmetry is broken.

Many details of such a scenario remain to be worked out, such as the
graceful exit from inflation and reheating of the Universe, and the
magnitude of the present-day dark energy and its possible relation to
supersymmetry breaking need to be understood better. Our framework offers
the possibility of tackling some of these issues in unconventional ways. In
the case of reheating, the dilaton mass during inflation, which is of
order $H$, appears insufficient to reheat the Universe. However, there may
be an alternative solution in our approach, using the supersymmetric
brane/D-particle model of~\cite{emw}. In addition to guaranteeing the
vanishing of the vacuum energy when the branes are static, the D-particles
of~\cite{emw} may cluster and form primordial black holes (PBHs) on the
branes and/or in the bulk space at the end of the inflationary period.

A possible scenario for the formation of such PBHs is the reduction of
the propagation velocity of the recoiling branes as they approach the
orientifold planes, where they eventually stop. Such a variation in the
recoil velocity would break the scale invariance of the spectrum of
primordial density perturbations, and it might be possible 
that $n_s$ in (\ref{inflq7}), which in that case 
would get modified  by higher order 
corrections involving (higher powers of) time dependent $\epsilon_i$, 
becomes greater than 1 at small scales,
in some analogy with curvaton effects in conventional cosmologies. 
This would lead to an increase of
the initial density contrast of the D-particles nearby, i.e., within
distances of order $\ell_s$, and on the D-brane world, and hence increase
the probability of forming small PBHs~\cite{carr}. For instance, these
might be formed by dust-like collapse of the D particles on the brane.  
The masses of the PBHs produced in this way would be controlled by the
scale at which the effective bump in the initial spectrum of perturbations
appears. In the above scenario, this would be during the last e-foldings
of inflation, when the adiabaticity of the relative motion of the branes
is violated. The subsequent Hawking evaporation of these black holes may
provide a source for radiation in the Universe and the required reheating
process, as advocated in the context of a hybrid inflationary scenario
in~\cite{linde2}.

There are many issues of course that should be carefully looked at in this
scenario for reheating, which we reserve for a future publication.
However, we are convinced that many of these and other issues related to
inflation and vacuum energy can only be understood within a stringy
framework, and hope that this paper may contribute to the formulation of
these problems within non-critical Liouville string theory.

This is a much broader framework than critical string theory, and allows
for the possibility of connecting various string vacua, including
metastable ones, that may not be possible otherwise. This is because of
the essential non-equilibrium nature of non-critical strings, which seems
particularly appropriate for discussing the early history of the Universe.  
The non-equilibrium physics at this epoch appears well suited to the use
of Liouville strings.

\section*{Acknowledgements}

We would like to thank H.~Hofer and F.~Pauss for their 
interest and support. 
N.E.M. wishes to thank the 
Department of Theoretical Physics of the University of Valencia 
(Spain) for hospitality during the
last stages of this work.
The work of
N.E.M. is partially supported by the European Union through contract
HPRN-CT-2000-00152. The work of D.V.N. is supported by D.O.E. grant
DE-FG03-95-ER-40917.

\end{document}